%% file: main.tex
\documentclass[a4paper]{PoS}

\input{defs.tex}

\title{FLAG: Lattice QCD Tests of the Standard Model and Foretaste for Beyond}

\ShortTitle{FLAG: Lattice QCD Tests of the Standard Model }

\author{\speaker{Anastassios Vladikas}
        \thanks{On behalf of the FLAG collaboration}\\
        INFN, Sezione di Tor Vergata,\\
	c/o Department of Physics,\\ University of Rome ``Tor Vergata",\\
	Via della Ricerca Scientifica 1,\\
	I-00133 Rome, Italy\\
       E-mail: \email{vladikas@roma2.infn.it}}

\abstract{After a short presentation of the FLAG collaboration, we review lattice results related to pion, $K$-, $D$- and $B$-meson physics with the aim of making them easily accessible to the particle-physics community. Only a selection of FLAG averages or estimates is presented. For light flavours, we present results on the form factor $f_+(0)$, arising in semileptonic $K \rightarrow \pi$  transition at zero momentum transfer, as well as the decay-constants $f_K,f_\pi$ and their ratio. The consequences of these results for the CKM matrix elements $|V_{us}|$ and $|V_{ud}|$ are discussed. For heavy flavours we focus on $D$- and $B$-meson decay constants and form factors, as well as the CKM matrix elements $|V_{cs}|$, $|V_{cd}|$ and $|V_{ub}|$. In addition we briefly cover the recent advances stemming from the calculation the $B_K$-parameters and touch upon related current results relevant to the Physics beyond the Standard Model, which will be the subject of the next FLAG edition.}

\FullConference{Flavor Physics \& CP Violation 2015\\
		May 25-29, 2015\\
		Nagoya, Japan}

\begin{document}

\input intro.tex
\input light.tex
\input charm.tex
\input bottom.tex

\input bk.tex

\section*{Acknowledgements}
I acknowledge the valuable contribution of my FLAG collaborators. Without their constant efforts this short review would not have been possible. Special thanks go to 
P.~Boyle, T.~Kaneko, S.~Simula who have produced a timely update of Sect.~\ref{sec:light}, Y.~Aoki, M.~Della Morte, C.-J.D.~Lin for the updated results in Sect.~\ref{sec:charm} and
P.~Dimopoulos, H.~Wittig, B.~Mawhinney for those of Sect.~\ref{sec:BK}. Discussions with S.~D\"urr, R.~Horsley, M.~Golterman and R.~Sommer are also gratefully acknowledged.


\end{document}

%% file: defs.tex
\definecolor{orange}{rgb}{1.0,.6,0}

\newcommand{\bda}{\begin{\displaymath}\begin{array}{rl}}
\newcommand{\eda}{\end{array}\end{displaymath}}
\newcommand{\be}{\begin{equation}}
\newcommand{\ee}{\end{equation}}
\newcommand{\bdm}{\begin{displaymath}}
\newcommand{\edm}{\end{displaymath}}
\newcommand{\bea}{\begin{eqnarray}}
\newcommand{\eea}{\end{eqnarray}}

\newcommand{\bi}{\begin{itemize}}
\newcommand{\ei}{\end{itemize}}

\newcommand{\beq}{\begin{equation}}
\newcommand{\eeq}{\end{equation}}

\newcommand{\msbar}{{\overline{{\rm MS}}}}

\def\mev{{\rm MeV}}
\def\gev{{\rm GeV}}
\def\tev{{\rm TeV}}

\newcommand{\bd}{\begin{displaymath}}
\newcommand{\ed}{\end{displaymath}}

\newcommand{\figurebox}[2]{\fbox{\vbox to#2in{\hbox to #1in{\hfil}\vfil}}}
\def\good{\raisebox{0.35mm}{{\color{green}$\bigstar$}}}
\def\bad{\raisebox{0.35mm}{\hspace{0.65mm}{\color{red}$\blacksquare$}}} 
\def\soso{\hspace{0.25mm}\raisebox{-0.2mm}{{\color{green}\Large$\circ$}}}

\def\figaver{\raisebox{0.35mm}{\hspace{0.65mm}{\color{black}$\blacksquare$}}} 
\def\figgood{\raisebox{0.35mm}{\hspace{0.65mm}{\color{green}$\blacksquare$}}} 
\def\figsoso{\raisebox{0.35mm}{\hspace{0.65mm}{\color{green}$\square$}}} 
\def\figbad{\raisebox{0.35mm}{\hspace{0.65mm}{\color{red}$\square$}}} 
\def\fignolat{\hspace{0.25mm}\raisebox{0.2mm}{{\color{blue}\Large\textbullet}}}

\def\ev{\mathrm{e\kern-0.1em V}}
\def\kev{\mathrm{ke\kern-0.1em V}}
\def\mev{\mathrm{Me\kern-0.1em V}}
\def\gev{\mathrm{Ge\kern-0.1em V}}
\def\tev{\mathrm{Te\kern-0.1em V}}

\def\n#1e#2n{{#1}\times 10^{#2}}

\def\bea{\begin{eqnarray}}
\def\eea{\end{eqnarray}}

\def\ods2{\mathcal{O}_{\Delta S=2}}
\def\zds2{Z_{\Delta S=2}}

\def\msbar{{\overline{\mathrm{MS}}}}

\makeatletter
\def\slash#1{{\mathpalette\c@ncel{#1}}} 
\def\big#1{{\hbox{$\left#1\vbox to1.012\ht\strutbox{}\right.\n@space$}}}
\def\Big#1{{\hbox{$\left#1\vbox to1.369\ht\strutbox{}\right.\n@space$}}}
\def\bigg#1{{\hbox{$\left#1\vbox to1.726\ht\strutbox{}\right.\n@space$}}}
\def\Bigg#1{{\hbox{$\left#1\vbox
to2.083\ht\strutbox{}\right.\n@space$}}}
\makeatother

%% file: intro.tex
\section{Introduction}
\label{sec:intro}

Lattice simulations performed by different groups involve different choices both at the level of formalism (lattice actions, number of sea flavours etc.) and at the level of resources (lattice volumes, quark masses etc.). Often this amounts to making different compromises which in turn introduce different systematic effects; thus not all lattice results of a given quantity are directly comparable. The Flavour Lattice Averaging Group (FLAG)  aims to answer, in a way which is readily accessible to non-experts, the question: When it comes to lattice quantiities of relevance to Flavour Physics, what is currently their ``best lattice value''? Two editions of the FLAG review have appeared so far~\cite{Colangelo:2010et,Aoki:2013ldr}, to which we will be referring as FLAG-1 and FLAG-2 for brevity. Currently the FLAG-3 report is under preparation
\footnote{For the composition of the collaboration, past and present, see the FLAG website:\\ \texttt{http://itpwiki.unibe.ch/flag}} 
and should appear in early 2016. 

In the present summary we will only show a subset of the results reviewed in FLAG-2~\cite{Aoki:2013ldr}, 
selecting those which are closely related to the topic of this conference (Flavour Physics \& CP-violation). The FLAG-2 compilation was based on results published in refereed journals by the 30th of November 2013, so they are somewhat outdated, while those of FLAG-3 is still in the works. As a compromise we show FLAG-2 results for $f_{D,B}$, the $D$- and $B$-meson semileptonic decay form factors, and the corresponding CKM matrix elements, since their update is still in a rather preparatory phase. We present {\bf preliminary} FLAG-3 results for $f_{\pi,K}$ (and also briefly for $f_D$), the $K$-meson semileptonic decay form factor, the corresponding CKM matrix elements and $B_K$, since their update is in an advanced stage. We also review two topics which go beyond the scope of FLAG: (i) an interesting analysis \cite{Bailey:2015wta} of the Standard Model (SM) prediction of $\epsilon_K$, based on the FLAG-2 estimate of $B_K$; (ii) the kaon oscillation $B$-parameters beyond the SM, which appeared in Refs.~\cite{Bertone:2012cu,Carrasco:2015pra}.

\begin{figure}[ht]
  \begin{center}
\hspace{-1.0cm}
      \includegraphics[width=10.0cm]{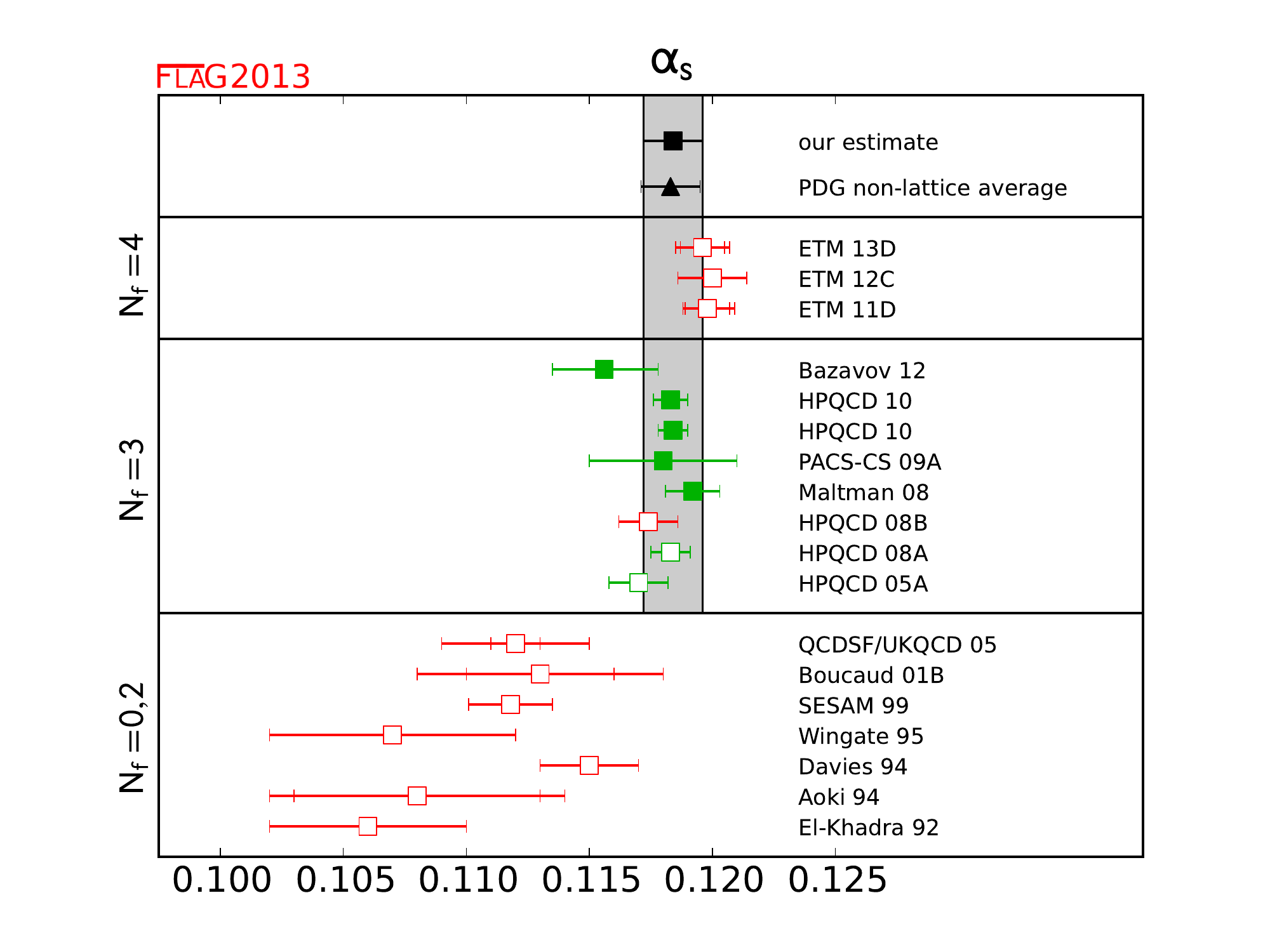}
   \end{center}
\vspace{-0.5cm}
\caption{ $\alpha_{\overline{\rm MS}}^{(5)}(M_Z)$, the coupling constant in the $\overline{\rm MS}$ scheme at the $Z$ mass.
For the reference labels see Ref.~\cite{Aoki:2013ldr}. The results labeled $N_f=0,2$ use estimates for $N_f=3$, obtained by first extrapolating in $N_f$ from $N_f=0,2$ results.
Since this is not a theoretically justified procedure, these are not included in our final estimate and are thus given a red symbol. The black triangle indicates the outcome of the PDG analysis excluding lattice results.}
\label{alphasMSbarZ}
\end{figure}

FLAG analyses are based on a critical review of lattice results. A number of criteria have been fixed, providing compact information on the quality of a computation. These are typically related to the quality of continuum and chiral extrapolations, finite volume effects, etc. The FLAG-2 rating is organized in Tables, in which different results are colour-coded as follows: \good~means that the relevant systematic error has been estimated in a satisfactory manner and is under control; \soso~means that a reasonable attempt at estimating the systematic error has been made, which however can be improved; \bad~means that there was no attempt or that the  attempt at controlling a systematic error was unsatisfactory. In the latter case the result is dropped from FLAG averages or estimates. The Tables are not presented here for lack of space. FLAG-3 adopted a similar rating, with somewhat differently phrased criteria. Several other issues are addressed in Ref.~\cite{Aoki:2013ldr}; e.g. how to average different results; how to arrive at a final estimate if an average is not possible; how to combine/correlate errors; how (not) to take conference proceedings into account. The interested reader should consult the introductory sections of Ref.~\cite{Aoki:2013ldr} for detailed explanations. For our purposes it suffices to keep in mind the following two FLAG policies:

Firstly, we recall that lattice simulations are carried out with a fixed number $N_f$ of dynamical (sea) quarks. So far results exist for $N_f = 0$ (quenched), $N_f = 2$ (two degenerate light quarks) $N_f = 2+1$ (including a heavier strange quark) and $N_f = 2+1+1$ (as before, plus charm). Quenched results are outdated and are thus omitted, with the exception of those related to $\alpha_s$; see Fig.~\ref{alphasMSbarZ}. FLAG averages or estimates are quoted separately for each $N_f$; results referring to different $N_f$ are never averaged, as they correspond to different field theoretic approximations of QCD.

Secondly, our Figures are organized as follows: results are clearly separated according to their $N_f$ value. Each datapoint is accompanied by an acronym, indicating the relevant reference, cited in Ref.~\cite{Aoki:2013ldr}. Lack of space prevents us from citing all these works here; however we do cite, whenever possible, the papers  containing the results on which FLAG averages/estimates are based. The plots are colour coded as follows:\\
\figaver~indicates a FLAG average or estimate; these results are also highlighted by a gray vertical band;\\
\figgood~refers to data that have passed our quality criteria and are used in the determination of the FLAG results (i.e. the points indicated by~\figaver); \\
\figsoso~indicates results with good control of the systematics (i.e. no red tags in the Tables of Ref.~\cite{Aoki:2013ldr}, which are left out of the average for some reason; e.g. not published in peer-reviewed journals, superseded by later results of the same collaboration, or simulations suffering from some uncontrolled effect which is not colour coded in the Tables)\footnote{In FLAG-3 this square is filled with ligh green; cf. Figs.~\ref{fig:lattice data},\ref{fig:VusVersusVud},\ref{fig:BK-3fig}.}\\
\figbad~refers to results that are not included in the average because they do not pass the rating criteria (i.e. they have at least one~\bad~in the Tables or do not pass for some other shortcoming); \\
\fignolat~indicates non lattice results, presented for comparison.

A good example of the effects of our rating in our final result is provided by the analysis of $\alpha_s$ in~Ref.~\cite{Aoki:2013ldr}. In Fig.~\ref{alphasMSbarZ} we show the strong coupling for 5 flavours, at the scale $M_Z$ in the $\msbar$ scheme. Preferring to err on the side of caution, the majority of FLAG-2 members opted for a conservative error estimate, based on Refs.~\cite{Maltman:2008bx, Aoki:2009tf,McNeile:2010ji,Bazavov:2012ka}, citing:
\begin{eqnarray}
  \alpha_{\overline{\rm MS}}^{(5)}(M_Z) = 0.1184(12) \,.
\label{eq:alpmz}
\end{eqnarray}
This is fully compatible with the PDG average, obtained by excluding lattice results: $\alpha_{\overline{\rm MS}}^{(5)}(M_Z) = 0.1183(12)$. Note that the error of the FLAG lattice estimate is as big as that of the above PDG non-lattice result.

%% file: light.tex
\section{Light Flavour Physics}
\label{sec:light}

We summarize state of the art lattice calculations of the leptonic kaon and pion decay constants and the kaon semileptonic decay form factor and provide an analysis in view of the Standard Model.  The pion decay constant is defined by the matrix element of the axial current $A_\mu = \bar d \gamma_\mu \gamma_5 u$ between vacuum and pion states:
\begin{equation}
\langle 0 \vert A_\mu \vert \pi^+(p) \rangle \,\, = \,\, i p_\mu f_{\pi^+}
\end{equation}
In this normalization, $f_{\pi^\pm} \approx 130$~MeV. Analogous expressions hold for $f_{K^+}$ and (in the following sections) the decay constants of charmed and bottom mesons. The kaon semileptonic decay form factor $f_+(q^2)$ represents one of the form factors relevant for the semileptonic decay $K^0 \to \pi^- l^+ \nu$, which depends on the momentum transfer (squared) $q^2$ between the two mesons. What matters here is its value at $q^2 = 0$.

All lattice results for the form factor $f_+(0$) and many available results for the ratio of decay constants have been computed in isospin-symmetric QCD. The reason for this unphysical parameter choice is that there are only few lattice simulations of SU(2) isospin-breaking effects. When combining lattice data with experimental results, we take into account the strong SU(2) isospin correction, either obtained in lattice calculations or estimated by using chiral perturbation theory, both for the kaon leptonic decay constant $f_{K^\pm}$ and for the ratio $f_{K^\pm}/f_{\pi^\pm}$. Here $f_{K^\pm}$ and $f_{\pi^\pm}$ are the isospin-broken decay constants in QCD (the electromagnetic effects have already been subtracted in the experimental analysis using chiral perturbation theory).
\begin{figure}[ht]
\hspace{-9mm}\includegraphics[height=6.8cm]{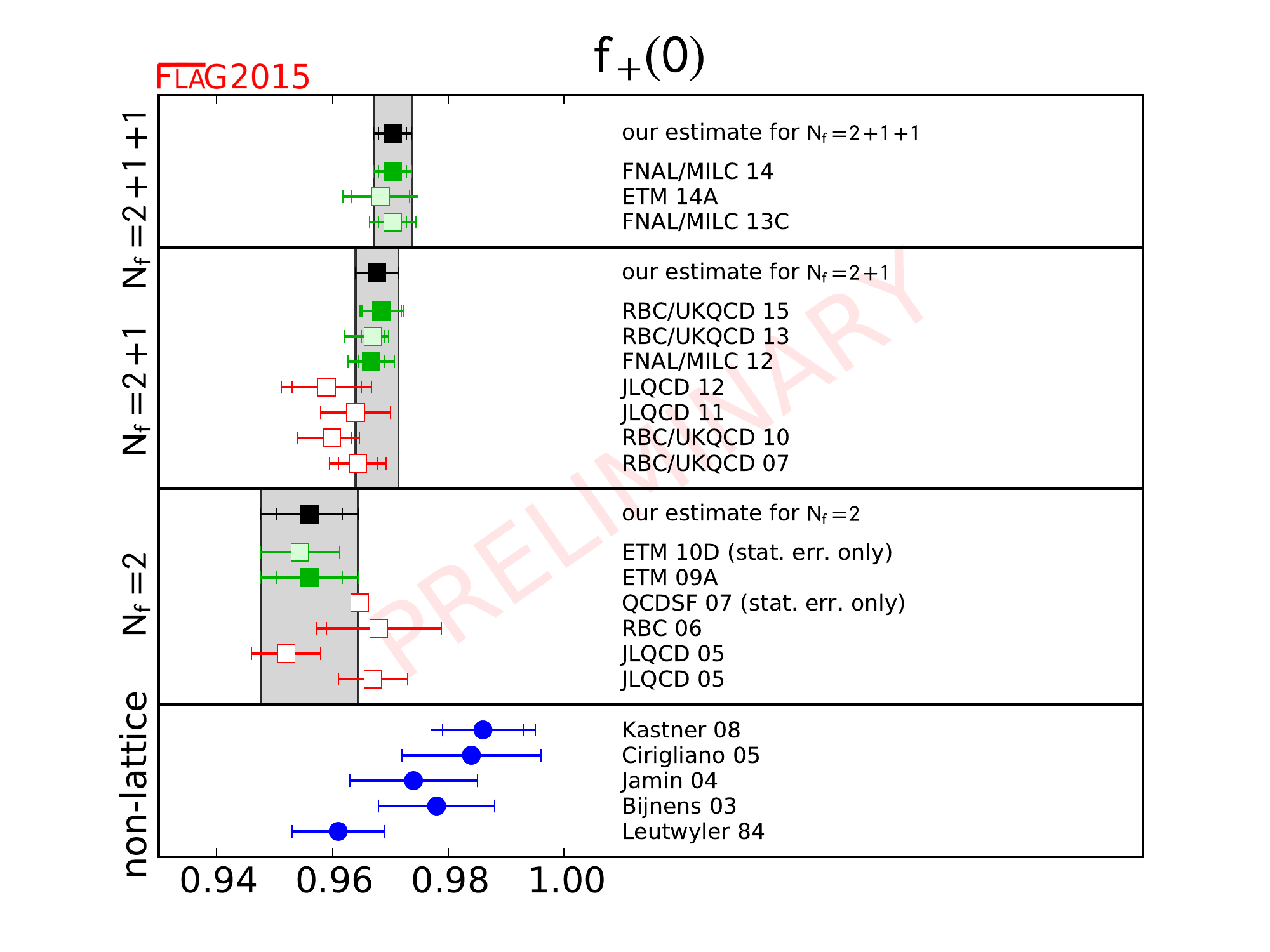}
\hspace{-1cm}
\includegraphics[height=6.8cm]{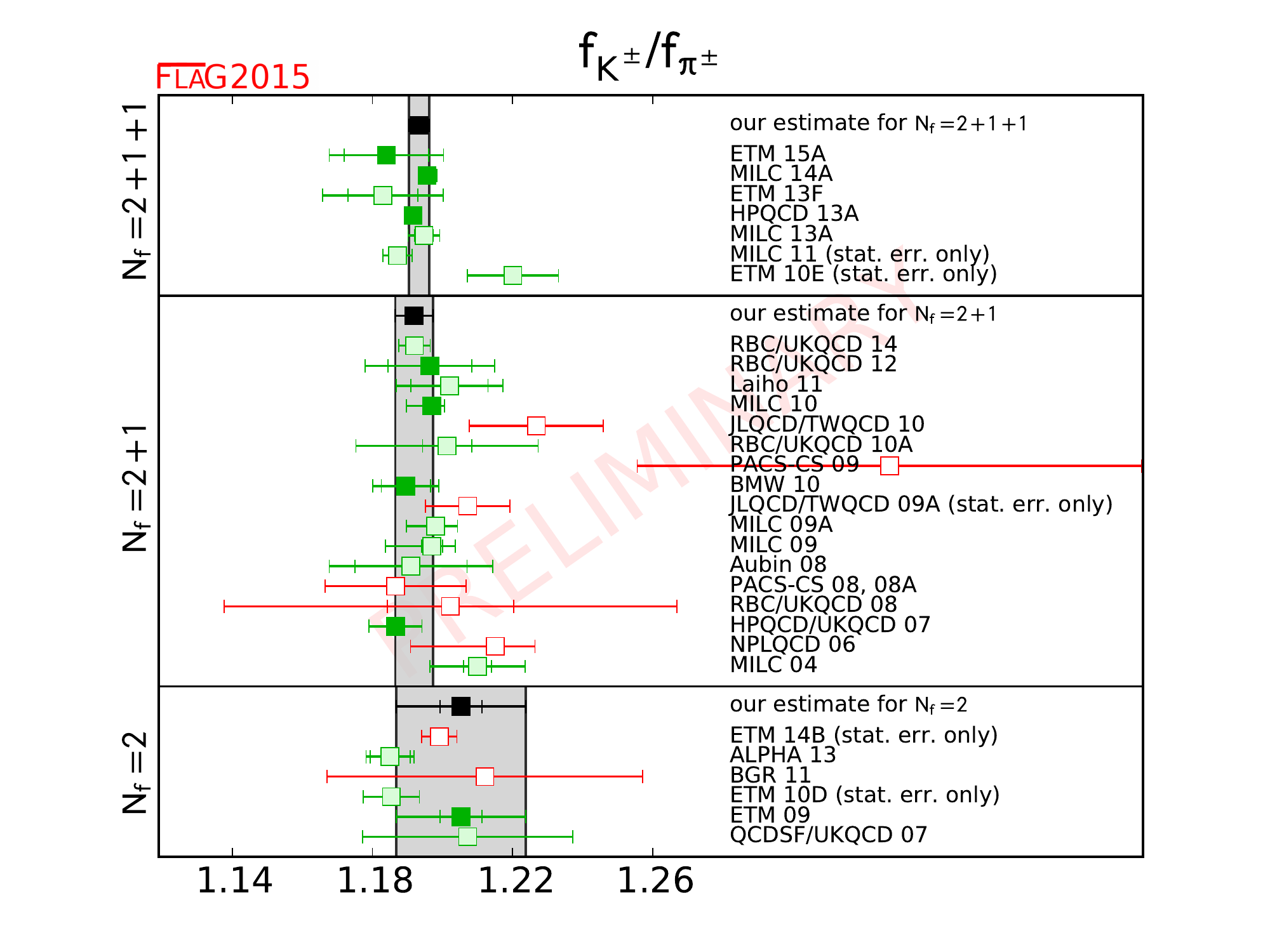}
\vspace{-0.5cm}
\caption{\label{fig:lattice data}Comparison of lattice results (squares) for $f_+(0)$ and $f_{K^\pm}/ f_{\pi^\pm}$ with various model estimates based on $\chi$PT (blue circles). The ratio $f_{K^\pm}/f_{\pi^\pm}$ is obtained in pure QCD including the SU(2) isospin breaking correction. For the reference labels see Ref.~\cite{Aoki:2013ldr}.}
\end{figure}
The plots in Figure~\ref{fig:lattice data} illustrate the {\bf preliminary} FLAG-3 compilation of data for $f_+(0)$ and $f_{K^\pm}/f_{\pi^\pm}$. The lattice data for the latter quantity are largely consistent even when comparing simulations with different $N_f$, while in the case of $f_+(0)$ a slight tendency to get higher values for increasing $N_f$ seems to be visible, not exceeding one standard deviation. 
The $N_f =2+1$ FLAG-average for $f_+(0)$ is based on FNAL/MILC 12~\cite{Bazavov:2012cd} and RBC/UKQCD 15~\cite{Boyle:2015hfa}, which we consider uncorrelated, while for $N_f = 2+1+1$ and $N_f = 2$ we consider directly the FNAL/MILC 14~\cite{Bazavov:2013maa} and ETM 09A~\cite{Lubicz:2009ht} results, respectively:
\begin{eqnarray}
    \label{eq:f+_direct_2p1p1}
    f_+(0) & = & 0.9704(24)(22)\,,     \hspace{1.0cm}(N_f=2+1+1)  \\
    \label{eq:f+_direct_2p1}
    f_+(0) & = & 0.9677(37) \,,           \hspace{1.7cm}(N_f=2+1)  \\
    \label{eq:f+_direct_2}
    f_+(0) & = & 0.9560(57)(62) \,.    \hspace{1.0cm}(N_f=2) 
\end{eqnarray}
The brackets in the first and third lines indicate the statistical and systematic errors respectively.
We stress that the results (\ref{eq:f+_direct_2p1p1}) and (\ref{eq:f+_direct_2p1}) include simulations with physical light quark masses. Note that the $N_f=2$ result remains unchanged from FLAG-2.

For the deacy constant ratio $f_{K^\pm}/f_{\pi^\pm}$ we quote:
\begin{eqnarray}
    \label{eq:fplus_direct_2p1p1}
    f_{K^\pm}/f_{\pi^\pm} & = & 1.193(3)\,,  \hspace{1.7cm}(N_f=2+1+1)  \\
    \label{eq:fplus_direct_2p1}
    f_{K^\pm}/f_{\pi^\pm} & = & 1.192(5) \,,  \hspace{1.7cm}(N_f=2+1)  \\
    \label{eq:fplus_direct_2}
    f_{K^\pm}/f_{\pi^\pm} & =  & 1.205(6)(17)  \,, \hspace{1.0cm}(N_f=2) 
\end{eqnarray}
where in the last row the first error is statistical and the second systematic. The first row is a preliminary update of FLAG-2, 
whereas the last two have remained unchanged. 

\begin{figure}[ht]
\hspace{-2mm}\includegraphics[width=8.5cm,height=7.5cm]{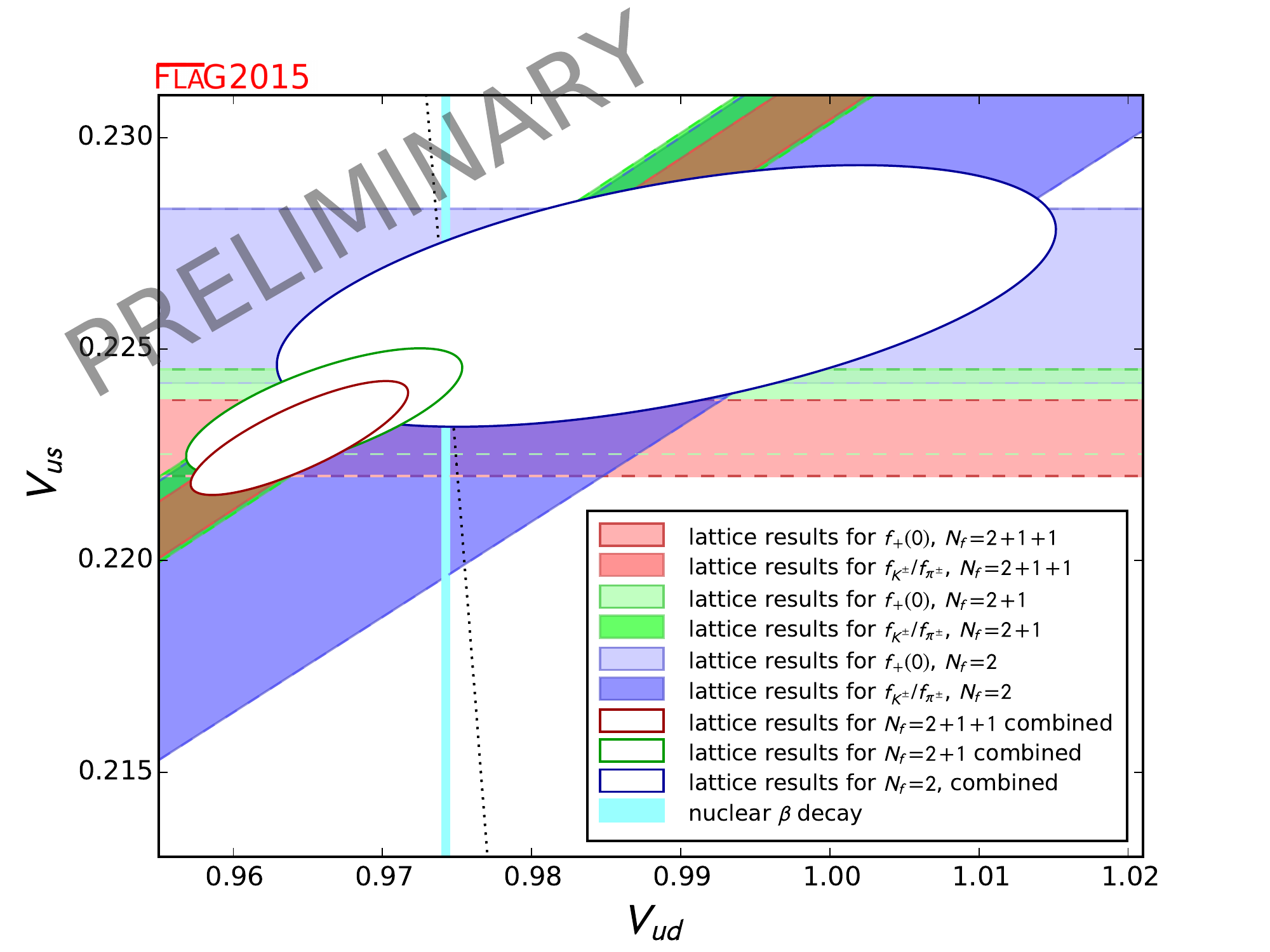}
\hspace{-1.0cm}
\includegraphics[width=8.5cm,height=7.5cm]{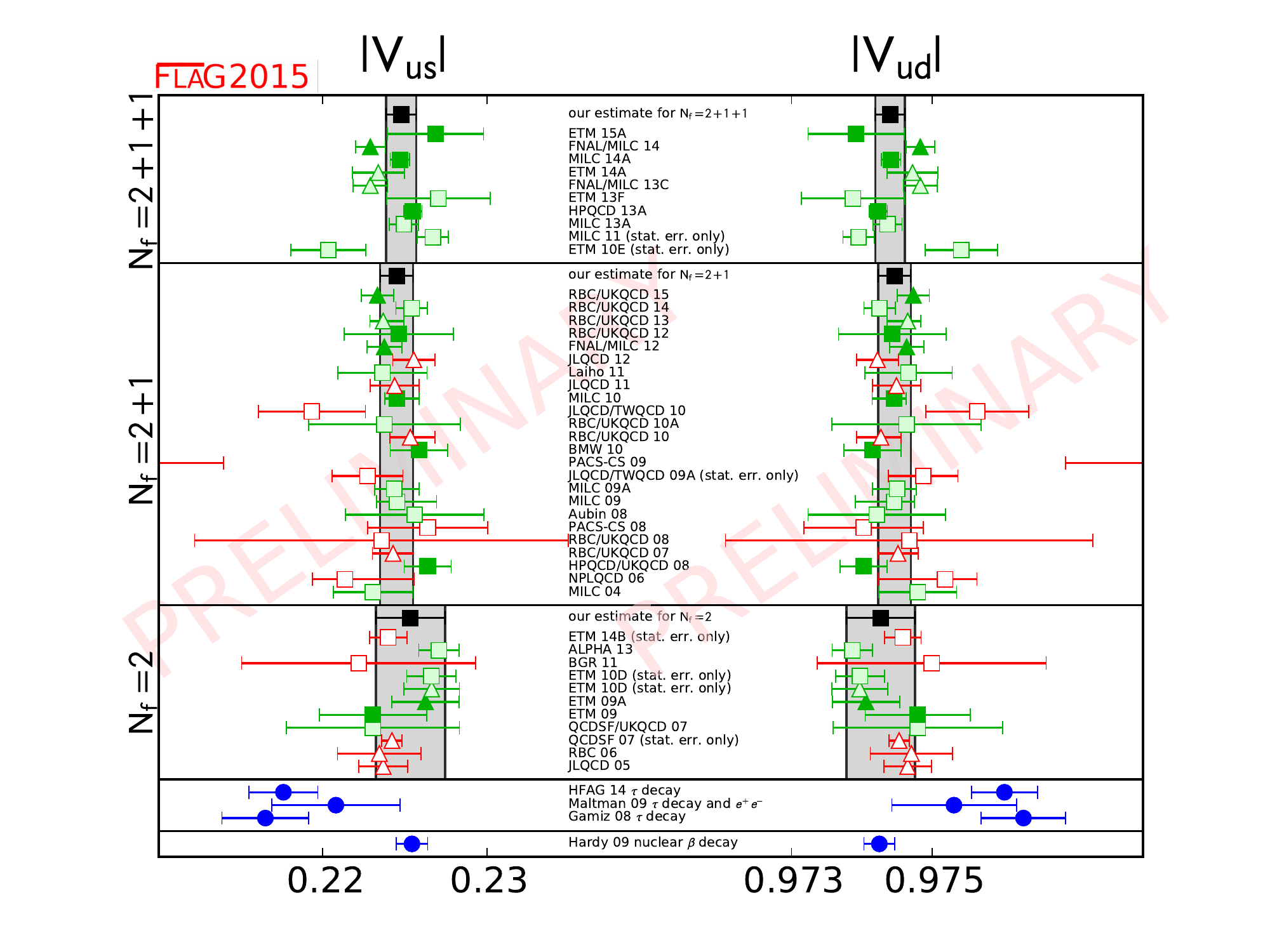}
\vspace{-0.5cm}
\caption{ {\bf Left:} The plot compares the information for $|V_{ud}|$, $|V_{us}|$ obtained on the lattice with the experimental result extracted from nuclear $\beta$ transitions. The dotted arc indicates the correlation between $|V_{ud}|$ and $|V_{us}|$ that follows if the three-flavour CKM-matrix is unitary.
{\bf Right:} Results for $|V_{us}|$ and $|V_{ud}|$ that follow from the lattice data for $f_+(0)$ (triangles) and $f_{K^\pm}/f_{\pi^\pm}$ (squares), on the basis of the assumption that the CKM matrix is unitary. 
The black squares and the grey bands represent our estimates, obtained by combining these two different ways of measuring $|V_{us}|$ and $|V_{ud}|$ on a lattice.
For comparison, the figure also indicates the results obtained if the data on nuclear $\beta$ decay and $\tau$ decay are analyzed within the Standard Model. For the reference labels see Ref.~\cite{Aoki:2013ldr}.}
\label{fig:VusVersusVud}
\end{figure}

Precision  experimental data on kaon decays very accurately determine the product $|V_{us}|f_+(0)$ and the ratio $|V_{us}/V_{ud}|f_{K^\pm}/f_{\pi^\pm}$ \cite{Antonelli:2010yf}: 
\begin{equation}
\label{eq:products}
|V_{us}| f_+(0) = 0.2163(5) \hspace{1cm} \;
\left|\frac{V_{us}}{V_{ud}}\right|\frac{ f_{K^\pm}}{ f_{\pi^\pm}} \;
=0.2758(5)
\end{equation}
Combining the above with our lattice estimates for $f_+(0)$ and $f_{K^\pm}/f_{\pi^\pm}$ we obtain the angles $|V_{ud}|$ and $|V_{us}|$, thus testing first row unitarity of the CKM matrix:
\begin{equation}
   \label{eq:CKM unitarity}
   |V_u|^2\equiv |V_{ud}|^2 + |V_{us}|^2 + |V_{ub}|^2 = 1
\end{equation} 
The tiny contribution from $|V_{ub}|$ is known much better than needed in the present context: $|V_{ub}|= 4.15 (49) \cdot 10^{-3}$ \cite{Beringer:1900zz}. 
Evidence for the validity of the relation (\ref{eq:CKM unitarity}) is shown in Figure~\ref{fig:VusVersusVud} (left). From our results for $f_+(0)$ and $f_{K^\pm}/f_{\pi^\pm}$ we obtain a corresponding range for the CKM matrix elements $|V_{ud}|$ and $|V_{us}|$, using the relations~(\ref{eq:products}). Consider first the results for $N_f = 2 + 1 + 1$. The range for $f_+(0)$ is mapped into the interval $|V_{us}| = 0.2229(9)$, depicted as a horizontal red band in Figure~\ref{fig:VusVersusVud} (left), while the one for $f_{K^\pm}/f_{\pi^\pm}$ is converted into $|V_{us}|/|V_{ud}| = 0.2311(7)$, shown as a tilted red band. The red ellipse is the intersection of these two bands and represents the 68\% likelihood contour, obtained by treating the above two results as independent measurements. Repeating the exercise for $N_f = 2 + 1$ and $N_f = 2$ leads to the green and blue ellipses, respectively.
The correlation between $|V_{ud}|$ and $|V_{us}|$ imposed by the unitarity of the CKM matrix is indicated by a dotted arc (in view of the uncertainty in $|V_{ub}|$, this is really a band of finite width, but the effect is too small to be seen here).
The plot shows that there is a slight tension with unitarity in the data for $N_f = 2 + 1 + 1$: Numerically, the outcome for the sum of the squares of the first row of the CKM matrix reads $|V_u|^2 = 0.980(10)$, which deviates from unity at the level of two standard deviations. 
In spite of this tension, it is fair to say that at this level the Standard Model passes a nontrivial test that exclusively involves lattice data and well-established kaon decay branching ratios. If we use the $\beta$ decay value of $|V_{ud}| = 0.97425(22)$ quoted in Ref.~\cite{Hardy:2008gy}, the test sharpens considerably: the lattice result (\ref{eq:f+_direct_2p1p1})) for $f_+(0)$ leads to $|V_u|^2 = 0.9989(8)$, while the one for $f_{K^\pm}/f_{\pi^\pm}$ (\ref{eq:fplus_direct_2p1p1}) implies $|V_u|^2 = 0.9999(7)$, thus confirming CKM unitarity at the permille level. The situation is similar for $N_f=2+1$: $|V_u|^2 = 0.983(12)$ from the lattice data alone. Using the $\beta$ decay value of $|V_{ud}|$ again the test sharpens, giving $|V_u|^2 = 0.9991(9)$ from the lattice value for $f_+(0)$ and $|V_u|^2 = 1.0000(6)$ from $f_{K^\pm}/f_{\pi^\pm}$. Repeating the analysis for $N_f = 2$, we find $|V_u|^2 = 1.029(35)$ with the lattice data alone, which is fully compatible with unity. When using the nuclear $\beta$ decay value of $|V_{ud}|$, we obtain $|V_u|^2=1.0004(10)$ from $f_+(0)$ and $|V_u|^2= 0.9989(16)$ from$f_{K^\pm}/f_{\pi^\pm}$.

The Standard Model implies that the CKM matrix is unitary. 
The precise experimental constraints quoted in (\ref{eq:products}) and the unitarity condition (\ref{eq:CKM unitarity}) then reduce the four quantities $|V_{ud}|$, $|V_{us}|$, $f_+(0)$, $f_{K^\pm}/f_{\pi^\pm}$ to a single unknown: any one of these determines the other three within narrow uncertainties.
Fig.~\ref{fig:VusVersusVud} (right) shows that the results obtained for $|V_{us}|$ and $|V_{ud}|$ from the data on $f_{K^\pm}/f_{\pi^\pm}$ (squares) are quite consistent with the determinations via $f_+(0)$ (triangles). 
In order to calculate the corresponding average values, we rely on those results that have passed the various rating criteria; for details see Ref.~\cite{Aoki:2013ldr} and the forthcoming FLAG-3 update.
The comparison shows that the lattice result for $|V_{ud}|$ not only agrees very well with the totally independent determination based on nuclear $\beta$ transitions, but is also remarkably precise. 
On the other hand, the values of $|V_{ud}|$ which follow from the $\tau$ decay data if the Standard Model is assumed to be valid, are not in good agreement with the lattice results. The disagreement is reduced considerably if the analysis is supplemented with experimental results on electroproduction. The {\bf preliminary} FLAG-3 results for $|V_{us}|$ and $|V_{ud}|$ (assuming unitarity of the CKM matrix) are
\begin{eqnarray}
    \label{eq:Vusud-prelim1}
    |V_{us}| & =  0.2248(9)\,, \qquad |V_{ud}| & = 0.97440(21) \,, \hspace{1.0cm} (N_f=2+1+1)  \\
    \label{eq:Vusud-prelim2}
    |V_{us}| & =  0.2245(10)\,, \qquad |V_{ud}| & =  0.97447(23) \,, \hspace{1.0cm} (N_f=2+1)  \\
    \label{eq:Vusud-prelim3}
    |V_{us}| & = 0.2253(21)\,, \qquad |V_{ud}| & = 0.97427(49) \,. \hspace{1.0cm} (N_f=2)
\end{eqnarray}

We close this section by quoting, without details, results for the decay contants $f_\pi^\pm$ and $f_K^\pm$. This is possible only for works which have not used $f_\pi$ for ``setting the scale'' (i.e. for the determination of the finite UV cutoff in the simulations). For the pion decay constant we have only the Flag-2~\cite{Aoki:2013ldr} estimate for $N_f = 2+1$;  $f_\pi^\pm = 130.2(1.4)$~MeV, to be compared with the PDG~\cite{Beringer:1900zz} value $f_\pi^\pm = 130.41(20)$~MeV. For $f_K^\pm$ we quote the FLAG-3 {\bf preliminary} values:
\begin{eqnarray}
   \label{eq:fK}
  f_{K^\pm} & = & 155.6 ~ (0.4) ~ \mbox{MeV} \qquad \qquad (N_f = 2 + 1 + 1), \\ \nonumber
  f_{K^\pm} & = & 155.9 ~ (0.9) ~ \mbox{MeV} \qquad \qquad (N_f = 2 + 1), \\ \nonumber
  f_{K^\pm} & = & 157.5 ~ (2.4) ~ \mbox{MeV} \qquad \qquad (N_f = 2) \, ,
 \end{eqnarray}
to be compared with the PDG~\cite{Beringer:1900zz} result $f_{K^\pm} = 156.2(7)$~MeV.

%% file: charm.tex
\section{Charm Physics}
\label{sec:charm}

Leptonic and semileptonic decays of charmed $D$- and $D_s$- mesons are sensitive probes of $c \rightarrow d$ and $c \rightarrow s$ quark flavour-changing transitions. They enable the determination of the CKM matrix elements $\vert V_{cd} \vert$ and $\vert V_{cs} \vert$ (within the Standard Model) and a precise test of the unitarity of the second row of the CKM matrix. Here we summarize the status of lattice-QCD calculations of the charmed leptonic decay constants  $f_{D_{(s)}}$ and semileptonic form factors $f_+^{D\pi}(0)$, as reported in FLAG-2~\cite{Aoki:2013ldr} and shown in Figure~\ref{fig:fD}. Our final averages are:
\begin{figure}[tb]
\hspace{-0.8cm}\includegraphics[width=0.58\linewidth]{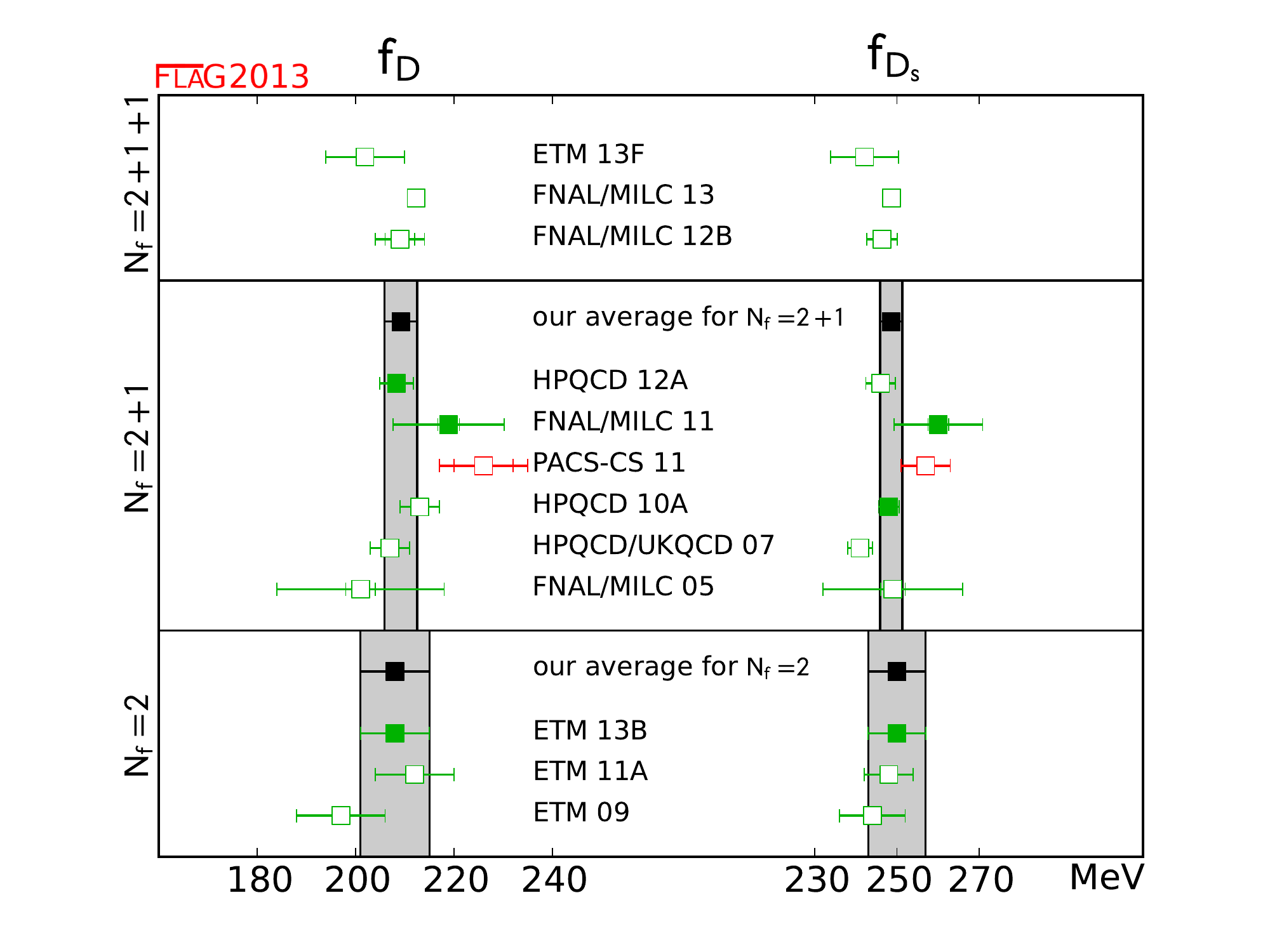} \hspace{-1cm}
\includegraphics[width=0.58\linewidth]{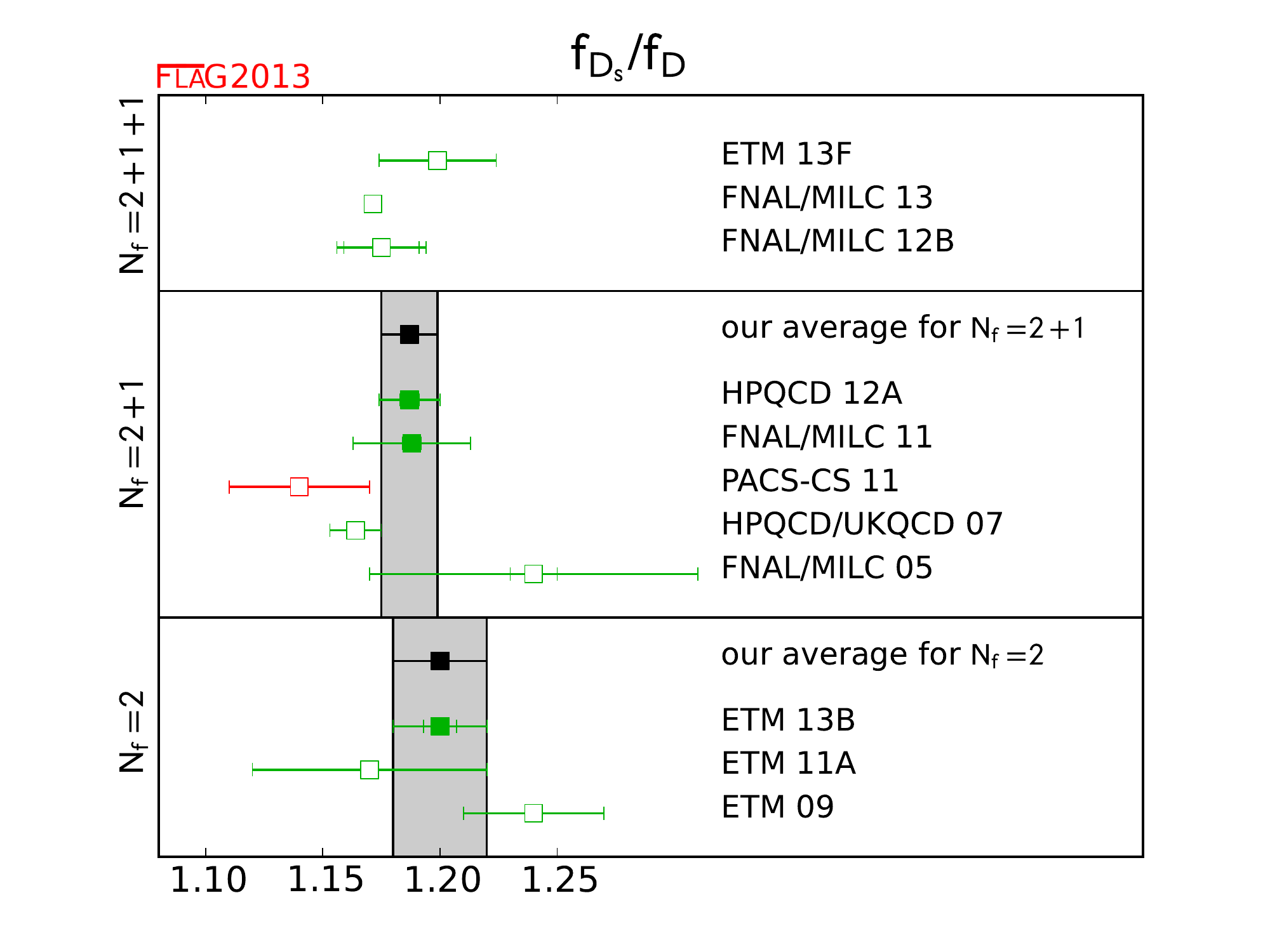}
\vspace{-0.5cm}
\caption{Decay constants of the $D$ and $D_s$ mesons. Errors in FNAL/MILC~13 are smaller than the
  symbols. For the reference labels see Ref.~\cite{Aoki:2013ldr}.}
\label{fig:fD}
\end{figure}
\begin{eqnarray}
& f_D = 209.2(3.3) \;{\rm MeV}, \,\, f_{D_s} =248.6(2.7) \;{\rm MeV}, & \,\, {{f_{D_s}}\over{f_D}}=1.187(12) \hspace{0.3cm} (N_f=2+1)
\label{eq:Nf3av} \\
& f_D = 208(7) \;{\rm MeV}, \,\, f_{D_s} = 250(7)\;{\rm MeV}, & \,\, {{f_{D_s}}\over{f_D}}=1.20(2) \quad \hspace{1.0cm} (N_f=2)
\label{eq:Nf2av}
\end{eqnarray}
The above $N_f=2$ is the result of the only reference in the literature at the time, ETM~13B~\cite{Carrasco:2013zta}. The $N_f=2+1$ estimates for $f_D$ and the SU(3) breaking ratio $f_{D_{s}}/f_D$ come from HPCQD~12A~\cite{Na:2012iu} and FNAL/MILC~11~\cite{Bazavov:2011aa} , while the result for $f_{D_{s}}$ comes from HPQCD~10A~\cite{Davies:2010ip} and FNAL/MILC~11~\cite{Bazavov:2011aa}. Because FNAL/MILC and HPQCD use a largely overlapping set of configurations, we treat the statistical errors as 100\% correlated.
A {\bf preliminary} FLAG-3 analysis gives $f_{D_{s}} = 249.8(2.3)$, upon including $\chi$QCD~14sea~\cite{Yang:2014sea}.
At the time of writing the FLAG-2 report the first $N_f=2+1+1$ computations had only appeared in conference proceedings, so no averages had been worked out.
Based on Refs.~~\cite{Bazavov:2014wgs} (FNAL/MILC~14wgs) and~\cite{Carrasco:2014poa} (ETM 14poa), a {\bf preliminary} analysis gives $f_D = 212.15(1.12)$~MeV, $f_{D_s} = 248.83(1.27)$~MeV and
$f_{D_s}/f_D = 1.1716(32)$.
We stress that since the accuracy of the lattice determinations of the
$D$ meson decay constant is rapidly improving, it will become important
in the future, especially when comparing to experimental numbers,
to distinguish between $f_{D^+}$ and the average of $f_{D^+}$ and $f_{D^0}$.

We now turn to the form factors for semileptonic $D\to \pi \ell\nu$ and $D\to
K \ell \nu$ decays. In practice, most lattice-QCD calculations focus on providing the vector form
factor at a single value of the momentum transfer, $f_+(q^2=0)$, which
is sufficient to obtain $|V_{cd}|$ and $|V_{cs}|$.  Because the decay
rate cannot be measured directly at zero momentum transfer, comparison
of these lattice-QCD results with experiment requires a slight
extrapolation of the experimental measurement. FLAG-2 did not quote any averages/estimates for  $N_f=2$ and $N_f = 2+1+1$, 
as at the time  they were at a preliminary stage. For $N_f = 2+1$ there was only one simulation satisfying all of our quality criteria (HPQCD~10B~\cite{Na:2010uf}), HPQCD~11~\cite{Na:2011mc}) from which we quote:
\begin{equation}
	f_+^{D\pi}(0) =  0.666(29) \,, \qquad f_+^{DK}(0) = 0.747(19) \qquad \qquad (N_f = 2+1)\,.
\label{eq:Nf2p1Dsemi}
\end{equation}

The results for the $D_{(s)}$ meson decay constants (related to the branching ratio for leptonic decays), as well as the semileptonic decays' form factors, lead to determinations of the CKM matrix elements $|V_{cd}|$ and $|V_{cs}|$ in the Standard Model.
For the leptonic decays, we use the latest experimental averages from
Rosner and Stone for the Particle Data Group~\cite{Rosner:2012np}
(where electromagnetic corrections of $\sim 1\%$ have been removed):
\begin{equation}
	f_D |V_{cd}| = 46.40(1.98)~{\rm MeV} \,, \qquad f_{D_s} |V_{cs}| = 253.1(5.3)~{\rm MeV} \,.
\end{equation}
We combine these with the average values of $f_D$ and $f_{D_s}$ in
Eqs.~(\ref{eq:Nf2av}) and~(\ref{eq:Nf3av}), to obtain $|V_{cd}|$ and $|V_{cs}|$.
For the semileptonic decays, we use the latest experimental averages
from the Heavy Flavour Averaging Group (HFAG)~\cite{Amhis:2012bh}.
\begin{equation}
	f_+^{D\pi}(0) |V_{cd}| = 0.146(3) \,, \qquad f_+^{DK}(0) |V_{cs}| = 0.728(5)  \,.
\end{equation}
and $f_+^{D\pi}(0)$ and $f_+^{DK}(0)$, of Eq.~(\ref{eq:Nf2p1Dsemi}), to obtain our preferred values for $|V_{cd}|$ and $|V_{cs}|$.
\begin{figure}[tb]
\hspace{-0.8cm}
\includegraphics[width=0.58\linewidth]{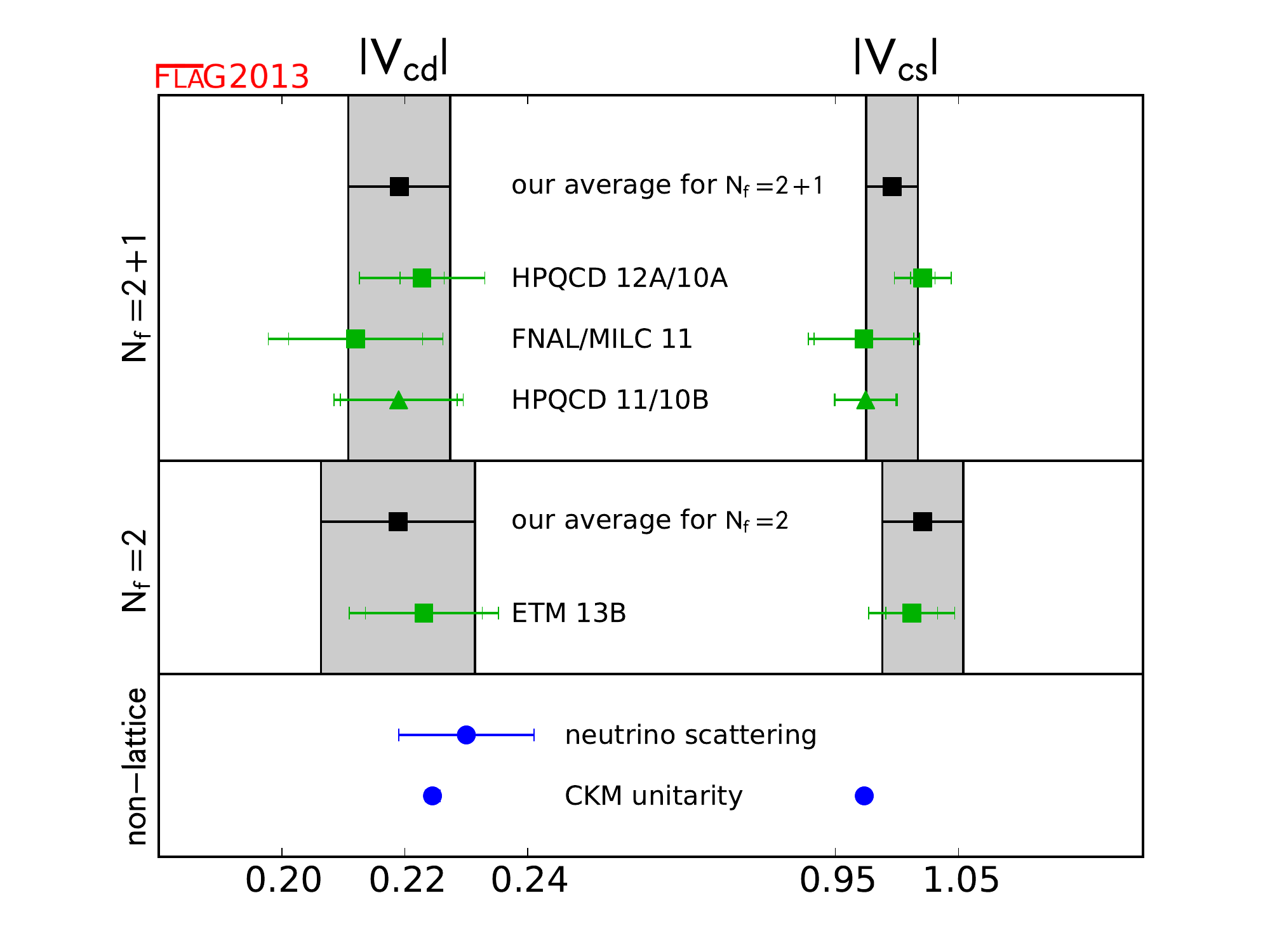} \hspace{-1cm}
\includegraphics[width=0.58\linewidth]{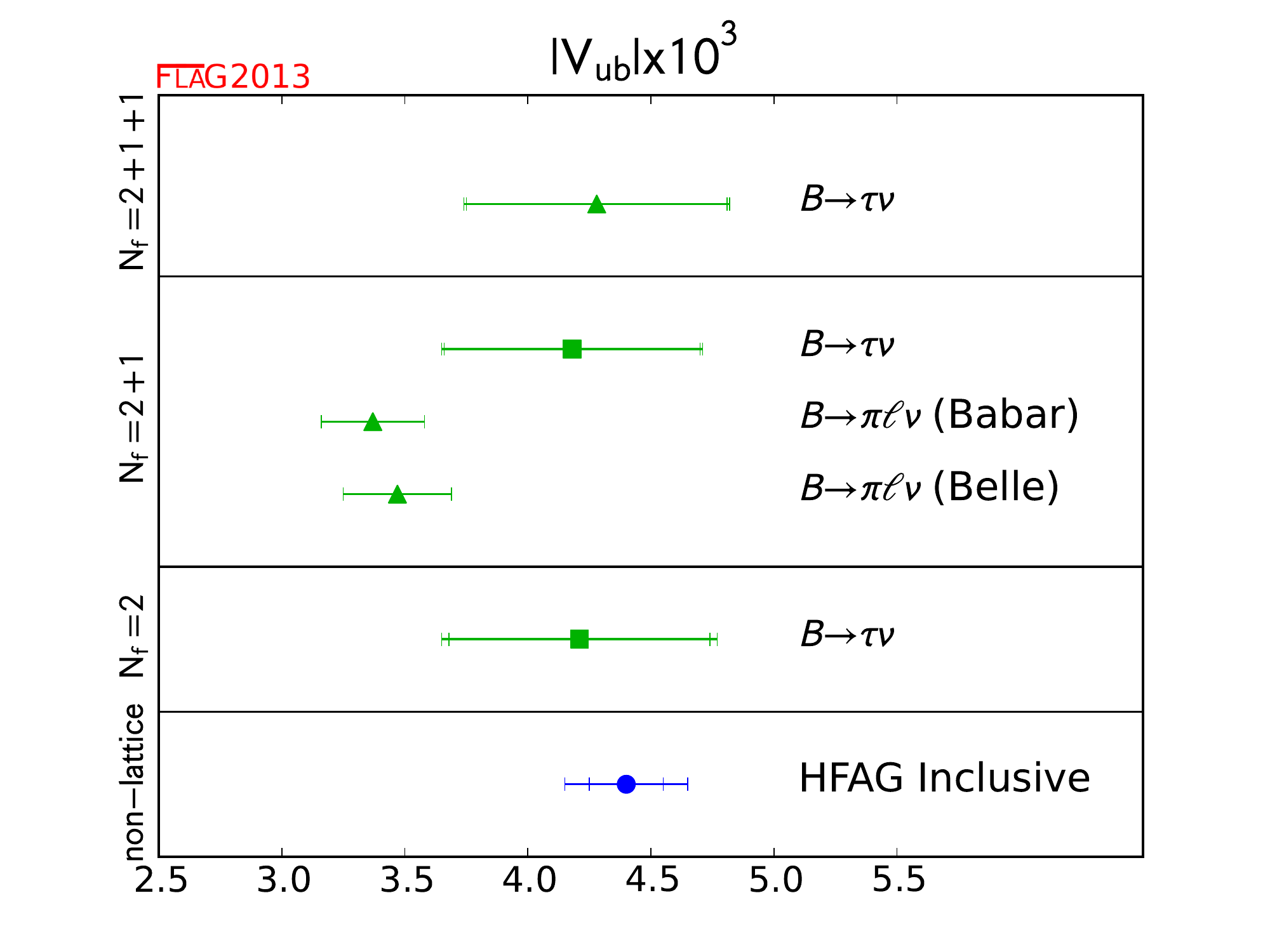}
\vspace{-0.5cm}
\caption{{\bf Left:} Comparison of determinations of $|V_{cd}|$ and $|V_{cs}|$
  obtained from lattice methods with non-lattice determinations and
  the Standard Model prediction based on CKM unitarity.  When two
  references are listed on a single row, the first corresponds to the
  lattice input for $|V_{cd}|$ and the second to that for $|V_{cs}|$.
  The results denoted by squares are from leptonic decays, while those
  denoted by triangles are from semileptonic
  decays. For the reference labels see Ref.~\cite{Aoki:2013ldr}.
  {\bf Right:} Comparison of determinations of $|V_{ub}|$ 
  obtained from lattice methods with non-lattice determinations based
  on inclusive semileptonic $B$ decays. The results
  denoted by squares are from leptonic decays, while those denoted by
  triangles are from semileptonic decays.}
\label{fig:VcdVcs-VubVcb}
\end{figure}
Fig.~\ref{fig:VcdVcs-VubVcb} displays the results for $|V_{cd}|$
and $|V_{cs}|$ from leptonic and semileptonic decays, and compares
them to determinations from neutrino scattering (for $|V_{cd}|$ only)~\cite{Beringer:1900zz}
and CKM unitarity~\cite{Rosner:2012np}. The determinations of $|V_{cd}|$ agree
within uncertainties.  The determination
of $|V_{cs}|$ from $N_f = 2+1$ lattice-QCD calculations of leptonic
decays is noticeably larger than that from both semileptonic decays
and CKM unitarity.  The disagreement between $|V_{cs}|$ from leptonic
and semileptonic decays is slight (only 1.2$\sigma$ assuming no
correlations), but the disagreement between $|V_{cs}|$ from leptonic
decays and CKM unitarity is larger at 1.9$\sigma$.  This tension is driven primarily by the HPQCD~10A~\cite{Davies:2010ip} calculation of
$f_{D_s}$, but we note that the
$N_f=2$ calculation of $f_{D_s}$ (ETM13B~\cite{Carrasco:2013zta})
leads
to the same high central value of $|V_{cs}|$, just with larger
uncertainties.

The $N_f=2+1$ averages for $|V_{cd}|$ and $|V_{cs}|$ in
Fig.~\ref{fig:VcdVcs-VubVcb} take correlations of different results into account.
Omitting details which the reader may find in Ref.~\cite{Aoki:2013ldr} , we simply state
that whenever there are correlations, they are assumed to be 100\%. 
We obtain
\begin{eqnarray}
	|V_{cd}| = 0.2191(83) \,, \quad |V_{cs}| = 0.996(21)  \,, \qquad  \qquad (N_f=2+1) \label{eq:VcdsNf2p1}
\end{eqnarray}
where the errors include both theoretical and experimental
uncertainties, and the error on $|V_{cs}|$ has been increased by
$\sqrt{\chi^2/{\rm dof}}=1.03$.

Using the determinations of $|V_{cd}|$ and $|V_{cs}|$ in
Eq.~(\ref{eq:VcdsNf2p1}), we can test the unitarity of the second row
of the CKM matrix.  We obtain
\begin{equation}
	|V_{cd}|^2 + |V_{cs}|^2 + |V_{cb}|^2 - 1 = 0.04(6) \,
\end{equation}
which agrees with the Standard Model at the percent level.  Given the
current level of precision, this result does not depend on the value
used for $|V_{cb}|$, which is of ${\mathcal{O}}(10^{-2})$.

%% file: bottom.tex
\section{Bottom Physics}
\label{sec:bottom}

Lattice-QCD calculations of $b$-quarks have an added complication not present for charm and light quarks: at the lattice spacings that are currently used in numerical simulations, the $b$-quark mass is of order one in lattice units. As a direct treatment of $b$-quarks with the fermion actions commonly used for light quarks would result in large cutoff effects, all current lattice-QCD calculations of $b$-quark quantities make use of effective field theory at some stage. The two most widely used general approaches are (i) direct application of effective field theory treatments such as HQET or NRQCD, which allow for a systematic expansion in $1/m_b$; or (ii) the interpretation of a relativistic quark action in a manner suitable for heavy quarks using an extended Symanzik improvement program to suppress cutoff errors. This introduces new problems (matching of HQET to QCD, renormalization, control of discretization effects...).

\begin{figure}[htb]
\hspace{-0.8cm}\includegraphics[width=0.58\linewidth]{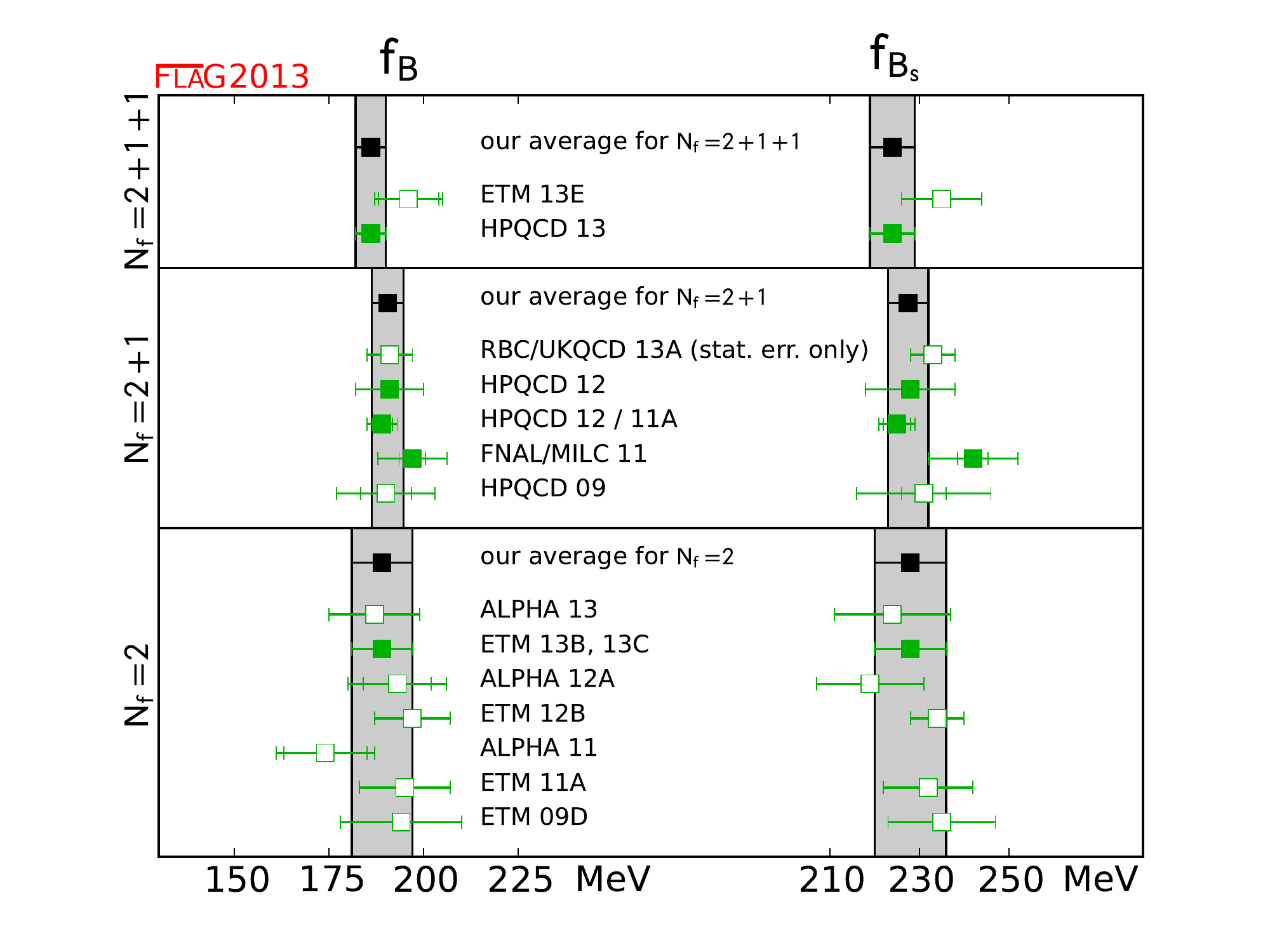}\hspace{-0.8cm}
\includegraphics[width=0.58\linewidth]{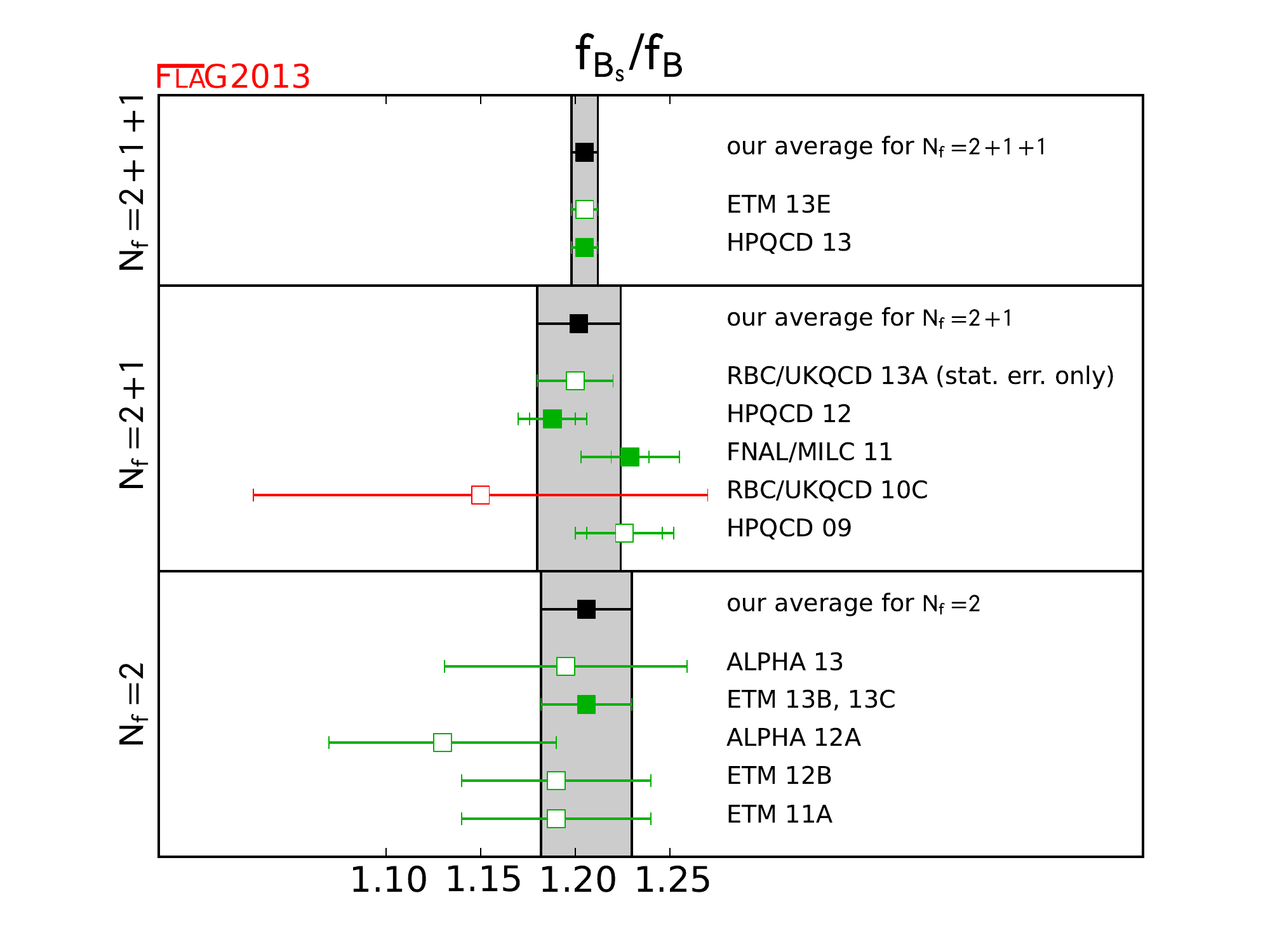}
\vspace{-2mm}
\caption{Decay constants of the $B$ and $B_s$ mesons. For the reference labels see Ref.~\cite{Aoki:2013ldr}.}
\label{fig:fB}
\end{figure}

In the present summary we will only quote results of the leptonic decay constants $f_B$ and $f_{B_s}$ and the CKM matrix element $|V_{ub}|$, once more leaving details and other results (on semileptonic form factors of heavy-to-light and heavy-to-charm decays, as well as neutral $B$-meson mixing matrix elements) to Ref.~\cite{Aoki:2013ldr}. The leptonic decay constants, shown in Fig.~\ref{fig:fB}, lead to the following FLAG-2 estimates:
\begin{eqnarray}
& f_B = 189(8) \;{\rm MeV}, \,\, f_{B_s} =228(8) \;{\rm MeV}, & \,\, {{f_{B_s}}\over{f_B}}=1.206(24) \hspace{0.5cm} (N_f=2)
\label{eq:fbav2} \\
& f_B = 190.5(4.2) \;{\rm MeV}, \,\, f_{B_s} =227.7(4.5) \;{\rm MeV}, & \,\, {{f_{B_s}}\over{f_B}}=1.202(22) \hspace{0.5cm} (N_f=2+1)
\label{eq:fbav21} \\
& f_B = 186(4) \;{\rm MeV}, \,\, f_{B_s} =224(5) \;{\rm MeV}, & \,\, {{f_{B_s}}\over{f_B}}=1.205(7) \hspace{0.7cm} (N_f=2+1+1)
\label{eq:fbav211}
\end{eqnarray}
Most results, obtained with degenerate light quarks, refer to average decay constants for $B^+$ and $B^0$. Some collaborations (FNAL/MILC, HPQCD) have started giving distinct results (they differ by about 2\%). As errors decrease with time, collaborations should start giving $B^+$ and $B^0$ results separately. The preliminary FLAG-3 update shows no significant changes for $N_f =2$ and $N_f=2+1$, while the $N_f = 2+1+1$ numbers remain unaltered.

We now use the lattice-determined Standard Model transition amplitudes for leptonic and semileptonic $B$-meson decays to obtain exclusive determinations of the CKM matrix element $|V_{ub}|$. The branching fraction for the decay $B\to\tau\nu_\tau$ has been measured by the Babar~\cite{Aubert:2009wt,Lees:2012ju} and Belle~\cite{Hara:2010dk,Adachi:2012mm} experiments. Combining the averaged data by the two experiments and the lattice FLAG2 averages, ref.~\cite{Aoki:2013ldr} quotes:
\begin{eqnarray}
|V_{ub}| = 4.21(53)(18) \times 10^{-3} & \qquad \qquad & N_f=2\,, \\
|V_{ub}| =  4.18(52)(9) \times 10^{-3} & \qquad \qquad & N_f=2+1\,,\\
|V_{ub}| =  4.28(53)(9) \times 10^{-3} & \qquad \qquad & N_f=2+1+1\,.
\end{eqnarray}

In semileptonic $B^0 \to \pi^- l^+ \nu$ decays we use as experimental input the Belle~\cite{Ha:2010rf}  and Babar~\cite{Lees:2012vv} resuts, combined with lattice form factor estimates for $N_f=2+1$, quoting:
\begin{eqnarray}
\mbox{global lattice + Babar:} & \qquad |V_{ub}| = 3.37(21) \times 10^{-3} \,,~~~~~~~N_f=2+1\,,\\
\mbox{global lattice + Belle:} & \qquad |V_{ub}| = 3.47(22) \times 10^{-3} \,,~~~~~~~N_f=2+1\,.
\end{eqnarray}
We do not quote a result for a combined lattice + Babar + Belle fit, since we are unable to properly take into account possible correlations between experimental results. 

Our results for $|V_{ub}|$ are summarized in Figure~\ref{fig:VcdVcs-VubVcb} (right), where
we also show the inclusive determinations from HFAG~\cite{Amhis:2012bh} for comparison.
The spread of values for $|V_{ub}|$ does not yield a clear picture.
We observe the well-known $\sim 3\sigma$ tension between
determinations of $|V_{ub}|$ from exclusive (lattice) and inclusive (HFAG) semileptonic
decays.  The determination of $|V_{ub}|$ from leptonic $B\to\tau\nu$
decay lies between the inclusive and exclusive determinations, but
the experimental errors in ${\rm BR}(B\to\tau\nu)$ are so large that
it agrees with both within $\sim 1.5\sigma$. The exclusive determination
of $|V_{ub}|$ will improve in the next few years with better
lattice-QCD calculations of the $B\to\pi\ell\nu$ form factor, while
the improvement in $|V_{ub}|$ from $B\to\tau\nu$ decays will have to
wait longer for the Belle~II experiment, which aims to begin running
in 2016.


%% file: bk.tex
\section{Neutral $K$-meson oscillations and $B_K$}
\label{sec:BK}

Indirect CP-violation in Kaons is measured by the parameter $\epsilon_K$. Followig Ref.~\cite{Bailey:2015wta} we express it here as
\begin{equation}
  \epsilon_K
  = e^{i\theta} \sqrt{2}\sin{\theta} 
  \Big( C_{\epsilon} \hat{B}_{K} X_{\rm {SD}} + \xi_{0} + \xi_{\rm{LD}} \Big) + \cdots
\label{eq:epsK}
\end{equation}
where some tiny correction terms have been dropped. This expression is valid in QCD with three light flavours; i.e. an effective theory, in which charm and heavier degrees of freedom have been integrated out. The factor
$ C_{\epsilon} = [G_{F}^{2} F_K^{2} m_{K^{0}} M_{W}^{2}]/[6\sqrt{2} \pi^{2} \Delta M_{K}]$ is a known quantity, and $\hat B_K$ is the  bag parameter, which incorporates low-energy QCD effects. It is defined through the hadronic matrix element of a four-fermion, dimension-six operator $Q^{\Delta S}_{\rm R}$, normalized by factors of the Kaon mass and decay constant:
\begin{equation}
B_{\rm K}(\mu)= \frac{{\left\langle\bar{K}^0\left| Q^{\Delta S=2}_{\rm R}(\mu)\right|K^0\right\rangle} }{{\frac{8}{3}f_K^2 m_K^2}}
\label{eq:BK}
\end{equation}
Note that $B_{\rm K}(\mu)$ is a renormalized quantity (thus the subscript in $Q^{\Delta S}_{\rm R}$) which depends on a chosen renormalization scheme and scale $\mu$. Its multiplication by a renormalization group evolution function, known in NLO perturbation theory, results in the scale independent (renormalization group invariant) quantity $\hat{B}_{K}$ which appears in eq.~(\ref{eq:epsK}). $B_{\rm K}(\mu)$ (and subsequently $\hat{B}_{K}$) is now known from several lattice computations, which have been reviewed by FLAG. The short-distance contribution is
\begin{equation}
X_{\rm SD} = \bar{\eta}\lambda^2 |{V_{cb}}|^2 \Bigg[ |{V_{cb}}|^2 (1-\bar{\rho}) \eta_{tt} S_0(x_t) (1 + r) + \left(1-\frac{\lambda^4}{8}\right) \left\{ \eta_{ct} S_0(x_c,x_t) - \eta_{cc} S_0(x_c) \right\} \Bigg]
\label{eq:XSD}
\end{equation}
where $S_0(x_{c,t})$ and $S_0(x_c,x_t)$ are the Inami-Lim functions (with $x_{c,t} \equiv m_{c,t}^2/M_W^2$) and $\eta_{tt}, \eta_{ct}, \eta_{cc}$ are factors known perturbatively to NLO, NNLO and NNLO respectively; the quantity $r$ is a known function of $S_0$'s and $\eta$'s. Long-distance effects $\xi_0$ from the absorptive part are estimated to have a $-7\%$ contribution~\cite{Blum:2011ng}, while long-distance effects $\xi_{\rm LD}$ from the dispersive part contribute only $2\%$~\cite{Christ:2012se} and will be neglected in the following discussion. For more details see Ref.~\cite{Bailey:2015wta} .

\begin{figure}[ht]
   \begin{center}
\hspace{-1.0cm}
      \includegraphics[width=15.0cm]{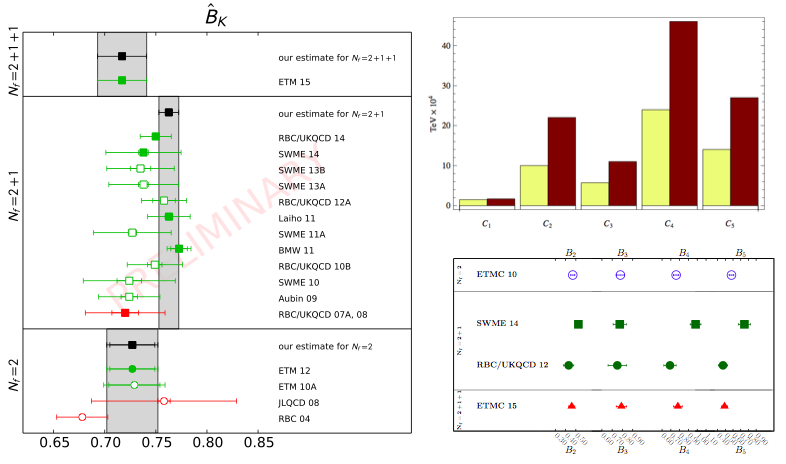}
   \end{center}
\vspace{-0.5cm}
\caption{{\bf Left:} Lattice results for the renormalisation group invariant $B_K$-parameter. For the reference labels see Ref.~\cite{Aoki:2013ldr}.
{\bf Right (top):} The lower bounds on the New Physics (NP) scale $\Lambda$, for generic NP flavor structure, obtained from the $N_f=2$ simulation of Ref.~\cite{Bertone:2012cu} (brown bars), compared to the bounds obtained form the quenched analysis of Ref.~\cite{Bona:2007vi} yellow bars). {\bf Right (bottom):} A compilation of $K^0$ meson bag-parameters $B_i, i = 2, . . . , 5$. From top to bottom data have been taken from Refs.~\cite{Bertone:2012cu,Jang:2014aea,Boyle:2012qb,Carrasco:2015pra}.}
\label{fig:BK-3fig}
\end{figure}

In Fig.~\ref{fig:BK-3fig} we present the {\bf preliminary} compilation of results of FLAG-3. For $N_f = 2+1$, the  result is\footnote{For comparison we also give $\hat{B}_{K} \,\, = \,\, 0.727(25)$ for $N_f = 2$ and $\hat{B}_{K} \,\, = \,\, 0.717(24)$ for $N_f = 2+1+1$.}
\begin{equation}
\hat{B}_{K} \,\, = \,\, 0.7627(97) \qquad \qquad N_f = 2+1
\end{equation}
This is compatible with the $N_f = 2+1$ value $\hat{B}_{K} = 0.7661(99)$ quoted in FLAG-2~\cite{Aoki:2013ldr}.

The FLAG-2 result has been used in ~\cite{Bailey:2015wta} for an interesting phenomenological analysis: Eqs.~(\ref{eq:epsK}) and (\ref{eq:XSD}) are used for a theoretical (Standard Model) prediction of $\epsilon_K$ and the value is compared to the known experimental result. For the Wolfenstein parameters $\bar \rho$ and $\bar \eta$ they prefer the so-called Angle-Only-Fit results of Ref.~\cite{Bevan:2013kaa} to the more standard UTfit / CKMfitter ones, because the latter contain unwanted dependence on $B_K$, $|V_{cb}|$ and $\epsilon_K$. They use $|V_{us}| \approx \lambda$ from $K_{\mu2}$ and $K_{l3}$ decays and $|V_{cb}| \approx A\lambda^2$. From Ref.~\cite{Alberti:2014yda} on inclusive $B \to X_c l \nu$ and $B \to X_s \gamma$ decays they take $|V_{cb}| = 42.21(78)  \times 10^{-3}$. From Ref.~\cite{Bailey:2014tva} on exclusive $B \to D^* l \nu$ decays they use $|V_{cb}| = 39.04(49)(53)(19) \times 10^{-3}$. These result to $|\epsilon_{\rm K}^{\rm SM}| = 1.58(18) \times 10^{-3}$ (exclusive) and $|\epsilon_{\rm K}^{\rm SM}| = 2.13(23) \times 10^{-3}$ (inclusive). Comparison with the experimental value $|\epsilon_{\rm K}^{\rm exp}| = 2.228(11) \times 10^{-3}$ shows a stress of $\Delta \epsilon_{\rm K} \equiv |\epsilon_{\rm K}^{\rm SM}|- |\epsilon_{\rm K}^{\rm exp}| = 3.6(2) \sigma$ in the exclusive case, which is absent in the inclusive one ($\Delta \epsilon_{\rm K} = 0.44(24) \sigma$). The former case requires further investigation.

A second interesting observation made in~\cite{Bailey:2015wta} concerns the error budget they report for $|\epsilon_{\rm K}^{\rm SM}|$, obtained with exclusive $|V_{cb}|$ and the FLAG-2 value of $\hat B_K$. Its error is dominated by $|V_{cb}|$ (nearly $41\%$), which is not surprising, given that $|V_{cb}|$ enters with a fourth power in Eq.~(\ref{eq:XSD}). In comparison, the contribution of the $\hat B_K$ error is a mere $1.6\%$, which indicates that lattice results nowadays are determined with errors which are clearly subdominant in certain phenomenological  analyses.

We close this section with a brief discussion of possible New Physics (NP) effects related to neutral Kaon oscillations, as recently analyzed in a model-independent way. A generalization of the effective $\Delta S = 2$ Hamiltonian is assumed, which contains operators beyond the one arising in the SM ; the amplitude is:
\begin{equation}
\langle \bar K^0| {\cal H}_{\rm eff}^{\Delta S = 2} | K^0 \rangle \,\, = \,\, C_1 \langle \bar K^0| O_1 | K^0 \rangle \,\, + \,\, \sum_{i=2}^5 C_i \langle \bar K^0| O_i | K^0 \rangle \,\, .
\end{equation}
The first term on the rhs is the SM contribution\footnote{i.e. $O_1 \equiv Q^{\Delta S=2}$ and $\langle \bar K^0 | O_1 | K^0 \rangle$ is the numerator of $B_K$ in Eq.~(\ref{eq:BK}).} and $O_{2,\cdots,5}$ are four-fermion operators of dimension-6, which parametrize NP effects. 
In analogy to $B_K \equiv B_1$, one defines  $B_{2, \cdots , 5}$.
The Wilson coedfficients $C_{2,\cdots, 5}$ are parametrized as $C_i = [F_i L_i]/\Lambda^2$, where $F_i$ stads for NP couplings, $L_i$ are coupling dependent loop factors and $\Lambda$ is the scale of NP. Assuming, for a generic strongly interacting theory with an unconstrained flavor structure, $F_i \sim L_i \sim {\cal O}(1)$, a generalized UT-fit analysis produces lower bounds for $\Lambda$; these depend very strongly (by several orders of magnitude!) on this assumption. To avoid accidental cancellations each contribution is analyzed separately. Fig.~\ref{fig:BK-3fig} (right, top) compares these lower bounds on $\Lambda$ resulting from two lattice simulations. They differ by the number of dynamical quark flavours in the sea, one being quenched ($N_f = 0$) and the other having $N_f=2$. More importantly, they also differ by the accuracy with which the $B_i$ parameters have been determined ($20\% - 23\%$ for $N_f = 0$ and $3\%-6\%$ for $N_f = 2$). The SM bound turns out to be several orders of magnitude weaker than those arising from NP operators.

This leads to the need of a more stringent determination of the NP $B$-parameters. Several groups are producing  results, which will be analyzed in FLAG-3. As a foretaste we show the recent comparison by Ref.~\cite{Carrasco:2015pra} in Fig.~\ref{fig:BK-3fig} (right, bottom). The situation is still unsettled, especially as far as $B_4$ and $B_5$ are concerned, in view of the different souces of systematic error characterizing the data.
